\documentclass[aps,reprint,groupedaddress,longbibliography]{revtex4-1}
\usepackage{graphicx}
\usepackage{amssymb}
\usepackage{amsmath}
\usepackage{setspace}

\usepackage[%
  colorlinks=true,
  urlcolor=blue,
  linkcolor=blue,
  citecolor=blue
]{hyperref}
\usepackage{url}

\begin{document}

\title{Robust inertial sensing with point-source atom interferometry for interferograms spanning a partial period}

\author{Yun-Jhih~Chen\textsuperscript{1,2}}
\altaffiliation{yunjhih.chen@colorado.edu}
\author{Azure~Hansen\textsuperscript{1}}
\author{Moshe~Shuker\textsuperscript{1}}
\author{Rodolphe~Boudot\textsuperscript{1,3}}
\author{John~Kitching\textsuperscript{1}}
\author{Elizabeth~A.~Donley\textsuperscript{1}}

\affiliation{\textsuperscript{1}Time and Frequency Division, National Institute of Standards and Technology, Boulder, Colorado 80305, USA}
\affiliation{\textsuperscript{2}Department of Physics, University of Colorado, Boulder, Colorado 80309, USA}
\affiliation{\textsuperscript{3}FEMTO-ST, CNRS, 26 rue de l'\'Epitaphe 25030 Besan\c{c}on, France}

\begin{abstract}
Point source atom interferometry (PSI) uses the velocity distribution in a cold atom cloud to simultaneously measure one axis of acceleration and two axes of rotation from the phase, orientation, and period of atomic interference fringe images. For practical applications in inertial sensing and precision measurement, it is important to be able to measure a wide range of system rotation rates, corresponding to interferograms with far less than one full interference fringe to very many fringes. The interferogram analysis techniques used previously for PSI are not sensitive to low rotation rates, which generates less one full interference fringe across the cloud, limiting the dynamic range of the instrument. We introduce an experimental method, new to atom interferometry and closely related to optical phase-shifting interferometry, that is effective in extracting rotation values from signals consisting of fractional fringes as well as many fringes without prior knowledge of the rotation rate. Our method uses four interferograms, each with a controlled Raman laser phase shift, to reconstruct the underlying atomic interferometer phase map.
\end{abstract}

\maketitle

\section{Introduction}\label{sec.intro}

Rotation sensing with optical and matter wave interferometers has a long tradition~\cite{Riehle.1991,Keith.1991}. Analogous to the Sagnac phase shift in a laser gyroscope, in a rotating frame, an atom interferometer experiences a Sagnac phase shift proportional to the inner product of the rotation vector and the Sagnac area~\cite{PathIntegral}. For guided-wave atom interferometers, the matter-wave trajectories follow the guide geometry, and the Sagnac area is fixed~\cite{Wu.2007,Moan.2020}. For free-space atom interferometers, the Sagnac area depends on atoms' initial velocity, and there are different approaches to handle this degree of freedom. For instance, in beam atom interferometers, the atomic beam has a wide velocity distribution in the longitudinal direction. At high rotation rates the signals from different velocity classes average out, leading to vanishing fringe contrast, whereas at low rotation rate the signals have high fringe contrast~\cite{Gustavson.1997}. In launched atom interferometers, cold-atom sources are used because they have velocity distribution with reduced width, and the atoms are launched with an initial velocity in a controllable fashion~\cite{Savoie.2018}.

Contrary to the above-mentioned conventional atom interferometry techniques, point source atom interferometry (PSI)~\cite{Dickerson.2013, Sugarbaker.2013} uses the velocity distribution of an expanding cold-atom cloud to simultaneously operate many Sagnac interferometers. In the expanding cloud, different velocity classes have different Sagnac areas, all of which contribute to the final interferometer signal. PSI maps the signal from each velocity class onto a unique point in the image plane, using the position-velocity correlation of the expanded point source. Because of that parallel operation, PSI measures one axis of acceleration and two axes of rotation from a single cold-atom source and without interleaving measurements. PSI is also distinct from conventional atom interferometer gyroscope techniques because a single rotation measurement with PSI yields a unique reading for the underlying rotation rate.

In PSI, a $\pi/2-\pi-\pi/2$ Raman laser pulse sequence is applied to an ensemble of cold atoms as it expands. Assuming the Raman laser beams are counter-propagating along the $z$-axis, the components of the rotation vector projected onto the $xy$-plane generate an interferometer phase gradient with the following $x$ and $y$ components,
\begin{equation} \label{eqn.kxky}
k_x = -\frac{2k_{\mathrm{eff}}T_R^2}{T_{\mathrm{ex}}}\Omega_y\ \mathrm{and}\
k_y = \frac{2k_{\mathrm{eff}}T_R^2}{T_{\mathrm{ex}}}\Omega_x
\end{equation}
where $k_{\mathrm{eff}}$ is the effective wave vector of the Raman laser, $\Omega_x$ and $\Omega_y$ are the components of the rotation vector in the $xy$-plane, $T_R$ is the time between the Raman pulses, and $T_{\mathrm{ex}}$ is the total expansion time of the cold-atom cloud ($T_{\mathrm{ex}}\ge 2T_R$)~\cite{Hoth.2016}. This phase gradient results in a sinusoidal fringe pattern in the population distribution across the final cloud. In principle, the rotation rate can be extracted from a single fringe image, but clockwise and counter-clockwise rotations remain indistinguishable. The rotation direction can be determined by taking a series of fringe images while scanning the Raman laser phase and observing the direction of fringe travel. The acceleration along the Raman beam direction and the Raman laser phase determine the phase of the fringes in the image.

Because PSI fringe images are essentially windowed pictures of monochromatic plane waves, a host of methods from wave optics can be applied to analyze the images. It is intuitive to use parametric fitting to extract the fringe contrast, orientation, frequency, and phase, where the latter three parameters can be converted to rotation and acceleration readings. However, parametric fitting to a fraction of a fringe period that has an unknown phase is challenging because, in practice, the plane wave is usually damped with a Gaussian envelope due to the spatial variation of the atomic distribution or contrast and there is other noise in the images. Since the phase gradient (and thus the number of fringes) is proportional to the rotation rate, a lower rotation rate that generates less than one fringe period can only be measured with significant error, except when the signal-to-noise ratio is excellent.

Dickerson~\textit{et al.} have described this problem of measuring small rotations in the partial-fringe regime (where there is less then one period of fringe in the images) when they measured the Earth's rotation using PSI~\cite{Dickerson.2013}. In their work, they applied a counter-rotation that nulls the phase gradient and thereby yields the Earth rotation rate. Instead of parametric fitting the entire cold-atom image to obtain the phase gradient as they scanned the counter-rotation, they detected the differential phase between the left-half and right-half of a cold-atom cloud with an ellipse fitting procedure~\cite{Foster.2002}.

However, the approach in \cite{Dickerson.2013} may be sensitive to fluctuations in the atomic density distribution. This concern is mentioned in the work by Sugarbaker~\textit{et al.}, where the authors used a different approach and demonstrated gyrocompassing \cite{Sugarbaker.2013}. In that study, an experimental phase shear, in the form of the Raman laser beam-tilt for the third Raman pulse, is applied to the cold-atom cloud to increase the number of fringes across the cloud to about 2.5 so that their parametric fitting procedure is operational.

We propose a different approach---instead of analyzing the fringe images, we convert the fringes in the population distribution into an interferometric phase map \cite{Hoth.2016}, which allows a direct analysis of the phase gradient and the acceleration phase, independent of the number of fringes. This is accomplished by an experimental method that converts four fringe images into a phase map, analogous to the phase shifting interferometry technique in optical interferometry~\cite{Bruning.1974,PSIBook1,PSIBook2}. We have termed the method ``Simple, High dynamic range, and Efficient Extraction of Phase map'', or ``SHEEP.'' The SHEEP method does not lose sensitivity in the partial-fringe regime.

As will be demonstrated in the following sections, the SHEEP method returns robust rotation readings and is independent of the fringe phase or the number of fringes in the image because the analysis is in the phase map domain. The SHEEP method does not require additional experimental means, such as compensation rotation or an additional Raman beam tilt, to perform well at low rotations. Further, the SHEEP method does not require a calibration of the fringe contrast. These features add up to unparalleled robustness and efficiency that makes the SHEEP method advantageous for real-time portable applications.

\begin{figure*}[!t]
\includegraphics[width=\textwidth]{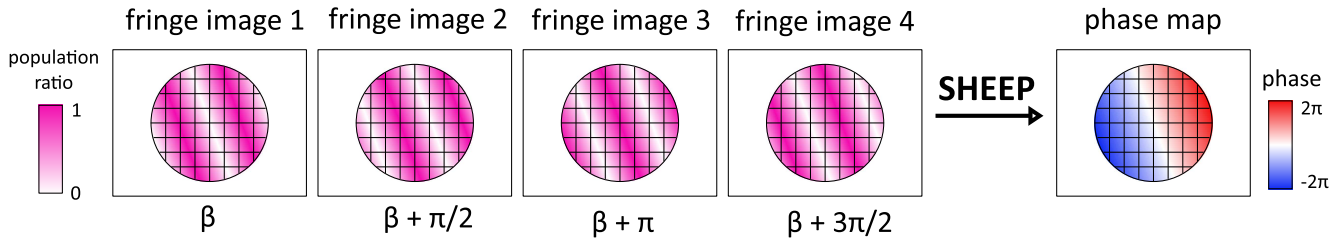}
\caption{Illustration of the SHEEP method. The SHEEP method takes a series of four images and each fringe image's phase differs from the previous one by $\pi/2$, where $\beta$ is an unknown fringe phase due to the acceleration phase and the Raman laser phase. Experimentally, the $\pi/2$ phase stepping is achieved by stepping the Raman laser chirp rate. The SHEEP method returns the phase map, which contains the values of $\beta$ and the rotation phase gradient. The operation is independent of the fringe contrast, fringe orientation, fringe frequency, and fringe phase. \label{fig.sheep}}
\end{figure*}

\section{Theoretical background}
\subsection{Extraction of the phase map}\label{sec.sheep}
Figure~\ref{fig.sheep} describes the concept of the SHEEP method which is to convert four fringe images into a phase map. This conversion is an inverse tangent function of a series of four fringe images, in which the phase of each fringe image differs from the previous one by $\pi/2$ \cite{GregThesis,Hoth.2016}. Experimentally, this stepping of the fringe phase by $\pi/2$ is done by increasing the Raman laser frequency chirp rate by $1/4T_R^2$ each time a new image is taken. Prior knowledge of fringe contrast, orientation, frequency, or phase is not needed to determine the phase map.

We label the four fringe images in population ratio as $P_1$, $P_2$, $P_3$, and $P_4$. The pixel value in each image is nominally described by a sinusoidal function of the phase,
\begin{equation}
P_L(x_i,y_j)=\frac{1}{2}\big[1-c(x_i,y_j)\cos(\phi(x_i,y_j)+2\pi\Delta\alpha_L T^2_R)\big],
\end{equation}
where $L =$ 1, 2, 3, or 4 is the image number, $(x_i,y_j)$ labels the pixel in the $i$th column and $j$th row, $c(x_i,y_j)$ is the contrast of the pixel $(x_i,y_j)$ and is not assumed to be constant over the range of the cold-atom cloud, $\phi(x_i,y_j) = \phi_\Omega(x_i,y_j)+\phi_a+\phi_l$ is the phase map that contains the rotation phase $\phi_\Omega(x_i,y_j)$, acceleration phase $\phi_a$, and the Raman laser phase $\phi_l$, and $\Delta\alpha_L$ is the difference in the Raman laser chirp rate from $\alpha_0$ in image $L$. $\alpha_0$ is the Raman laser chirp rate that compensates for the Doppler shift due to the free fall of the atoms. Without loss of generality, we assume $\Delta\alpha_1 = 0$.

We step the chirp rate such that $(\Delta\alpha_{L+1}-\Delta\alpha_L)\times2\pi T_R^2 =\pi/2$ and calculate the phase map as
\begin{equation}
\phi(x_i,y_j) =\tan^{-1} \Bigg[\frac{P_2(x_i,y_j)-P_4(x_i,y_j)}{P_3(x_i,y_j)-P_1(x_i,y_j)}\Bigg]+n\pi,\label{eqn.atan}
\end{equation}
where $n$ is an integer that removes discontinuities in the phase map. The mathematical form in Equation~\ref{eqn.atan} is similar to that of the phase-shifting interferometry in optical interferometry \cite{Bruning.1974,PSIBook1,PSIBook2}.

The contrast of each pixel $c(x_i,y_j)$ is a common factor of the numerator and denominator in the argument of the inverse tangent and cancels out in the calculation. Thus, the phase map is separated from the spatial distribution of atoms and the spatially-dependent Raman laser Rabi frequency. In other words, the phase of each pixel in the phase map is computed from the pixel's own amplitudes as it steps through a full 2$\pi$ oscillation. This operation is independent of the values of neighboring pixels, and thus robust against any background structure of the image.

By definition, the output range of the inverse tangent is between $-\pi/2$ and $\pi/2$, resulting in discontinuities in the phase map when the rotation rate is high enough that multiple fringes are created across the cloud. Unwrapping the inverse tangent output, i.e., determining the value of $n$ in Equation~\ref{eqn.atan}, does not change the phase gradient. We note that the value of $n$ changes the global offset of the entire phase map in steps of $\pi$, arising from the fact that the acceleration measurement in PSI is fundamentally ambiguous over phase multiples of 2$\pi$, as in conventional atom interferometers. Additional techniques can be implemented to resolve the ambiguity problem, which is especially important for portable applications \cite{Freier.2016,Bidel.2013,Wu.2019,yankelev.2020}.

\subsection{Phase map analysis versus fringe image analysis}\label{sec.FitModel}

The experimentally-obtained phase map $f(x,y)$ can be described by a simple linear model with only three fitting parameters,
\begin{equation}\label{eqn.plane}
f(x,y) =k_xx+k_yy+\phi_0,
\end{equation}
where $\phi_0$ is the sum of the acceleration phase and the Raman laser phase, $k_x$ is the rotation phase gradient in the $x$-axis, and $k_y$ is the rotation phase gradient in the $y$-axis. All three parameters are useful in inertial sensing. The rotation components are recovered from the phase gradients via the relations in Equation~\ref{eqn.kxky}. The acceleration along the $z$-axis, $a_z$, can be recovered by the relation $a_z = \phi_0/(k_{\mathrm{eff}}T_R^2)$, but this is not emphasized in the present work.

The experimentally-obtained fringe images can be modeled as a plane wave that has five fitting parameters,
\begin{equation}\label{eqn.planewave}
g(x,y) =c\cos(k_xx+k_yy+\phi_0)+g_0,
\end{equation}
where the additional fitting parameters are $c$, the contrast of the fringes (assumed to be constant over the range of the cold-atom cloud), and $g_0$, the overall background. If the Gaussian envelope cannot be removed by a normalized detection scheme, the fringe images may be modeled by including an additional Gaussian envelope, though at the cost of requiring more fitting parameters.

The phase gradients, the contrast, and the fringe phase are all fitting parameters in the sinusoidal function. In principle, the fitting procedure should not require prior knowledge to return reasonable values of the parameters. However, experimentally, parametric fitting of the fringe images with sinusoidal functions is not as robust as linear fitting of phase maps, as will be shown in Section~\ref{sec.ExpResults}. The simplification from sinusoidal fitting with five parameters (fringe analysis) to linear fitting with three parameters (phase map analysis) contributes to the superior robustness of the SHEEP method.

\section{Experimental setup and methods}\label{sec.setup}

The experimental setup has been described in our previous work~\cite{gyro.2019}; however, in the present work, we use a different experimental timing sequence to acquire the four fringe images for demonstrating the SHEEP method (without changing the hardware).

The experiment uses $^{87}$Rb atoms laser-cooled in a glass vacuum cell with a 1~cm$^2$ cross-sectional area. At the beginning of each experimental sequence, the atoms are loaded into a six-beam magneto-optical trap (MOT). The cloud of atoms is made smaller and colder with compressed-MOT and molasses stages, resulting in an initial atomic cloud with a diameter of 0.4~mm and a temperature lower than 10~$\mu$K. The freely-expanding cloud of atoms is then transferred to the $F = 1$, $m_F = 0$ sublevel of the ground state by optical pumping. The Raman $\pi/2-\pi-\pi/2$ laser pulses are applied in the direction of the local gravitational field. The $\pi$-pulse duration is $5~\mu$s and the time between pulses is $T_R=7.8$~ms. After the pulse sequence, the cloud is imaged in the plane transverse to the Raman laser beams with state-selective absorption imaging. At the time of imaging, $T_{\mathrm{ex}}=25.9$~ms, the cloud has expanded by a factor of 4 and fallen by 3 mm. We perform normalized detection by taking an image of the atoms in the $F = 2$ state, repumping all atoms to $F = 2$, and then taking a second image of the atoms in the $F = 2$ state. The fringe image is obtained by taking the ratio of the first image to the second image and plotting the population ratio of the atoms in $F = 2$ state. The total experimental cycle time is 166.7 ms. Due to our camera frame rate, it takes one second to acquire the four population ratio images to establish one phase map.

The laser beams that drive the two-photon Raman transitions between the $^{87}$Rb $F = 1$ and $F = 2$ states are spatially superimposed with orthogonal circular polarizations. A bias magnetic field is applied in the direction parallel to the laser beams. The laser beam driving the transition from the $F = 1$ state to an intermediate state is denoted as the ``F1 beam,''  and the other laser beam driving the transition from the $F = 2$ to the intermediate state is denoted as the ``F2 beam.'' The F1 beam is retro-reflected back to the atoms after passing through the cell, and the polarization is reversed from $\sigma^-$-polarization to $\sigma^+$-polarization. The counter-propagating pair of the F1 beam and the F2 beam is used to transfer photon momenta to the atoms. The co-propagating pair of the F1 beam and the F2 beam drives magnetically-sensitive transitions, which are tuned off resonance by the bias magnetic field.

During the Raman interrogation, the frequency of the F1 beam is chirped with a direct digital synthesizer at $\alpha_0=-25.1$~kHz/ms to compensate for the Doppler shift due to the free fall of the atoms. Varying the chirp rate around $\alpha_0$ also scans the overall interferometer phase and translates the spatial fringes across the image plane.

Rotation of the lab frame causes the direction of both Raman beams to rotate about the free-falling atoms. We simulate rotation by piezoelectrically sweeping the angle of the retro-reflection mirror of the Raman F1 beam, which causes a phase shift that are equivalent to the shift caused by the rotation of the lab frame for small mirror rotation angles \cite{Lan.2012}.

\section{Experimental results} \label{sec.ExpResults}

Our objective is to compare the phase map analysis with the more typical fringe analysis by parametric fitting. For the phase map analysis, we extract the phase map with the SHEEP method, then apply a 2D linear fit to the phase map with Equation~\ref{eqn.plane}. For the fringe analysis, we process all four images and then take the average of the results since we convert a series of four fringes images into one phase map. The full cycle period for the measurements with the two techniques are equal. We first perform principal component analysis (PCA) to the set of four images to remove some background noise, and then apply a 2D curve fit to the four cleaned fringe images with the plane wave function in Equation~\ref{eqn.planewave}. We refer to this approach as the ``PCA method,'' which has been used in previous PSI works~\cite{Dickerson.2013,Sugarbaker.2013,gyro.2019}. PCA is model-free, and has been used in different atomic systems to remove noises in the images before analyzing the data with a physical model~\cite{Segal.2010,Dubessy.2014} \footnote{We note that there are other widely-used model-free noise-removal techniques, such as the one demonstrated in~\cite{Ockeloen.2010}. However, considering the overall laser excitation and probing scheme used in our work (that the background noise does not just come from one probe laser beam), and the moving nature of the interferometric fringe patterns, PCA is more appropriate for our application.}. It should be noted that the comparison is between one phase map and four fringe images. A sample of the phase map and the fringe images can be found in the Appendix.

\subsection{Azimuthal scans of fixed-magnitude rotation vectors}\label{sec.TimeTrace}
\begin{figure*}[!t]
\includegraphics[width=\textwidth]{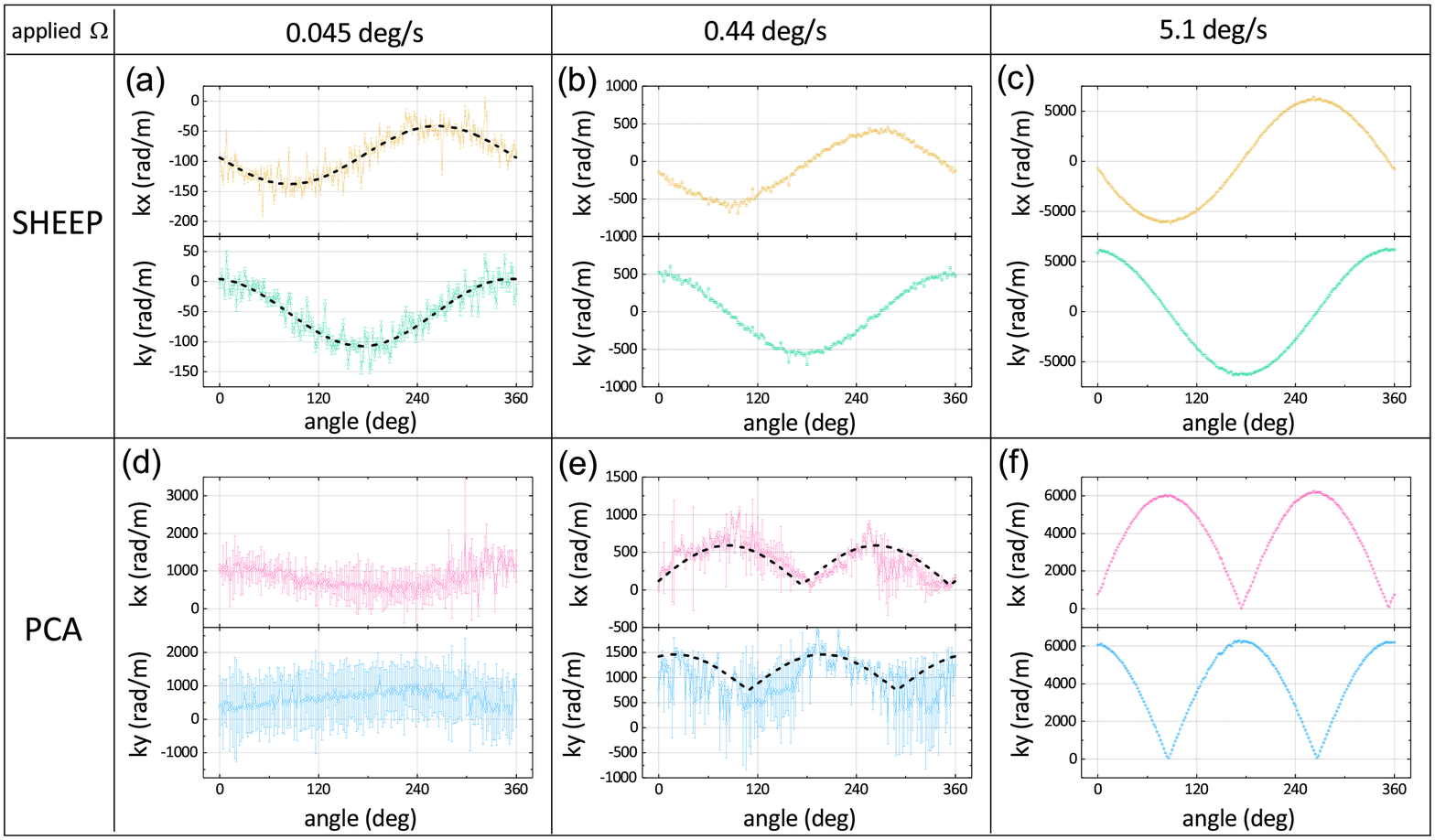}
\caption{Azimuthal scans of fixed-magnitude rotation vectors, with the experimental data processed with the SHEEP method and the PCA method, for the indicated rates. The half length of the error bar attached to each point is the value of the 2D fit error. In (a), the dashed lines are sine and cosine fits to the time traces. In (e), the dashed lines are magnitude-sine and cosine fits. An example of the fringe images for the case of 5.1 deg/s rotation ((c) and (f)) can be found in the Appendix. $T_R = 7.8$~ms for all three scans. \label{fig.TimeTrace}}
\end{figure*}

We apply a rotation vector in the plane perpendicular to the Raman laser beams with a fixed rotation rate while stepping the direction of the rotation from 0 to 360 degrees in two-degree steps. In each rotation step, we record a series of four fringe images with fringe phases of $0$, $\pi/2$, $\pi$, and $3\pi/2$ to the unknown acceleration phase and the Raman laser phase. We process the sets of four images with the SHEEP method and the PCA method separately and plot the measured rotation phase gradients versus the azimuthal angle of the applied rotation vector.

The traces obtained by the SHEEP method and the PCA method are shown in Fig.~\ref{fig.TimeTrace} for three different rotation rates. Because the rotation vector precesses about the direction of the Raman laser beams, the $x$-component of the phase gradient follows a sine curve while the $y$-component of the phase gradient follows a cosine curve, via the relation in Equation~\ref{eqn.kxky}.

Qualitatively, the SHEEP method resolves the sinusoidal variation in all three cases, while the PCA method loses sensitivity as the rotation rate becomes smaller.

In Fig.~\ref{fig.TimeTrace} the traces of the PCA method are plotted as the absolute value of the rotation phase gradient because the sign of the phase gradients $k_x$ and $k_y$ returned by the plane wave fitting is meaningless. An extended experimental procedure and additional analysis would be needed to determine the direction of the fringe travel, and thus the sign of the phase gradient. This is not relevant for the sensitivity comparison here; however, the ability of the SHEEP method to distinguish between a rotation and a counter-rotation further highlights its practical advantages.

We note that there is a non-zero phase gradient offset even when there is no applied rotation. This phase gradient offset was slowly changing over the timescale of the measurements presented in this work. An example of this phase gradient offset is visible in Fig.~\ref{fig.TimeTrace} (a), where both sinusoidal curves are centered at a non-zero value in the vertical direction. The non-zero, slow-varying phase gradient offset in the experimental data may arise from technical sources, such as the Raman laser beam wavefront not being uniform.

\subsection{Statistical error of single rotation measurement}\label{sec.FitError}
\begin{figure*}[!t]
\includegraphics[width=5in]{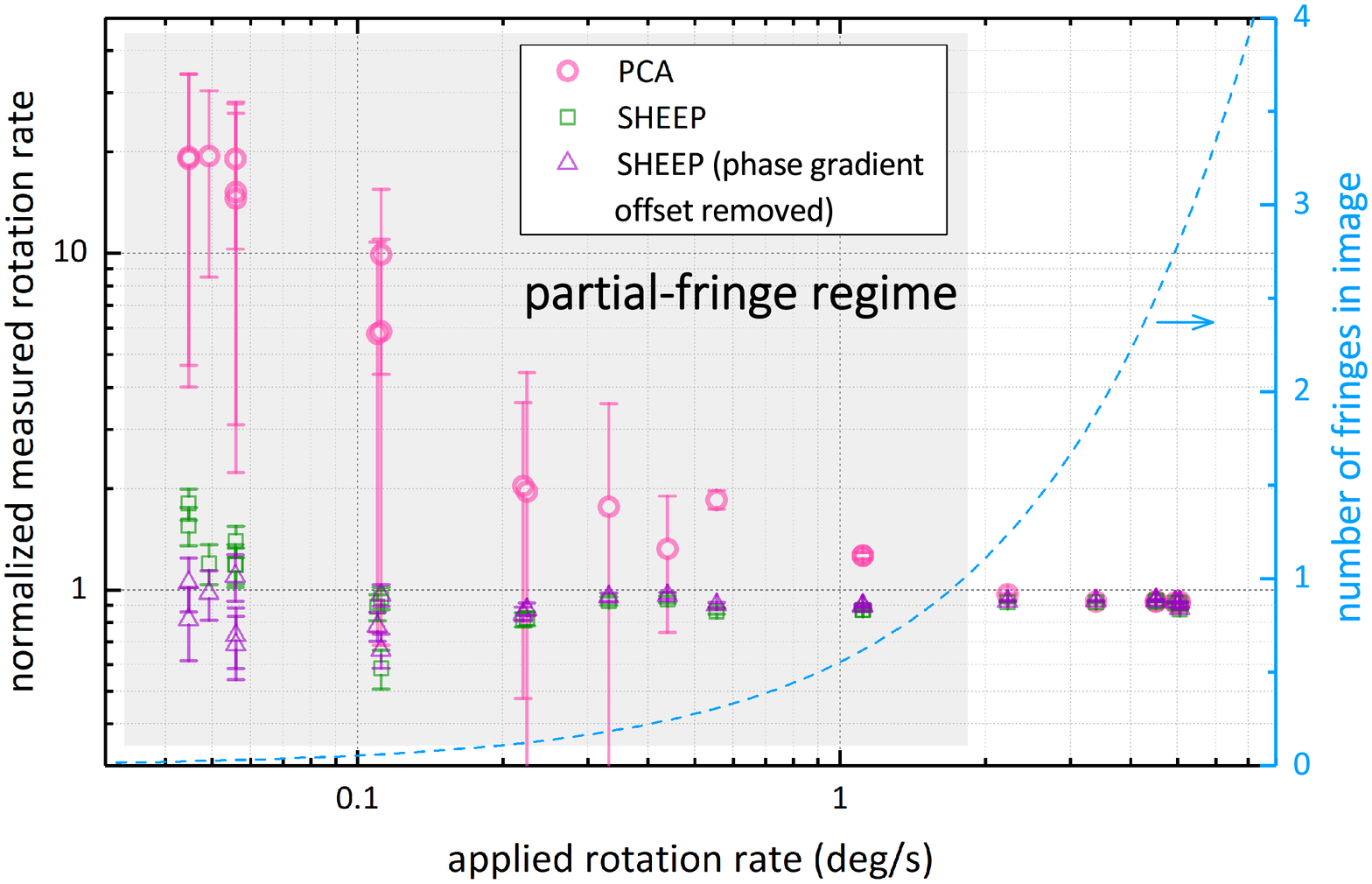}
\caption{Fit results of single rotation measurement, in both the SHEEP method and the PCA method, versus the applied rotation rate. The left vertical axis is the measured value returned from the fitting procedure and is plotted as normalized values. The half length of the error bar is the normalized 2D-fit standard error. The blue dashed line (vertical axis on the right) shows the calculated number of fringes along the diagonal of an image versus rotation rate for an ideal point source with $T_R = 7.8$~ms and $T_\mathrm{ex} = 25.9$~ms. \label{fig.FitError}}
\end{figure*}

Figure \ref{fig.FitError} shows the results of single rotation measurement (of a set of four images) returned from the PCA method (pink circles) and the SHEEP method (green squares and purple triangles) at different rotation rates. Both the measured rotation value and the error in Fig.~\ref{fig.FitError} are normalized to the applied rotation rate. The two methods return similar normalized values and errors when the images have more than one fringe; however, the PCA method worsens immediately when the number of the fringes in the image is less than one, while the normalized value and the error of the SHEEP method remains low down to the lowest rotation rates studied. In other words, the two methods are similar in the multiple-fringe regime; however, in the partial-fringe regime, the PCA method gradually loses sensitivity. Contrary to the PCA method, the SHEEP method remains robust without any correction and the dynamic range is at least 10 dB better than that of the PCA method. The dynamic range of the SHEEP method is further improved when the phase gradient offset (at zero applied rotation rate) is removed (purple triangles).

For a perfect point source, the normalized rotation should be one, i.e., with a scale factor that equals one. However, the scale factor is expected to be around 0.9 in this work, because the cold-atom cloud expands by a factor of about four and is therefore not an ideal point source. Avinadav and Yankelev \textit{et al.} have demonstrated methods to correct this effect \cite{Avinadav.2000}.

The rotation phase gradient, and thus the number of fringes, scales with the rotation rate and the time between the Raman laser pulses (assuming $T_{\mathrm{ex}}\approx 2T_R$; see Equation~\ref{eqn.kxky}). In this work, $T_R$  is $7.8$~ms due to the small dimension of our experimental setup. We also plot the calculated number of fringes of an ideal point source in Fig.~\ref{fig.FitError} in order to highlight the fact that the measurement covers both the multiple-fringe regime and partial fringe regime. Due to the reduced scale factor, the actual number of fringes is slightly lower~\cite{Hoth.2016}.

\subsection{Allan deviation of the rotation measurement at one second}\label{sec.AllanDev}
\begin{figure*}[!t]
\includegraphics[width=5in]{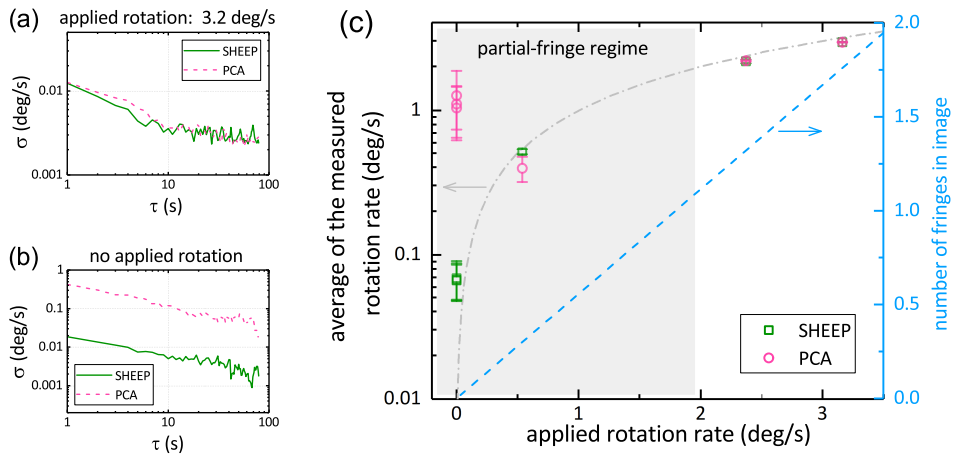}
\caption{Allan deviation plots of rotation measurements and measured values versus applied rotation rates. (a) and (b) are plots of the Allan deviation curves of rotation measurements at an applied rotation rate of 3.2~deg/s and zero applied rotation rate, respectively. We fit the curves from 1 to 10 seconds with the function $A/\sqrt{\tau}$, where the fit parameter $A$ is an estimate of the Allan deviation at one second. We plotted the measured rotation rates versus the applied rotation rate in (c). The half-lengths of the error bars equal the respective coefficients $A$. Note that the rightmost data points (squares and circles) are overlapped. The dot-dashed line equals the applied rate (for a scale factor of one). The dashed line with the vertical axis on the right shows the number of fringes in the image (calculated in the same manner as described in the caption of Fig.~\ref{fig.FitError}). \label{fig.AllanDev}}
\end{figure*}

We also record a time series of images where the applied rotation rate and the rotation direction are fixed. We repeat the measurement at a few different applied rotation rates, including zero applied rotation rate. Figures~\ref{fig.AllanDev} (a) and (b) show two samples of Allan deviation plots at an applied rotation rate of 3.2~deg/s and zero applied rotation rate, respectively. The Allan deviation curves from the SHEEP method and the PCA method are not significantly different from each other in (a). However, in the case of zero applied rotation rate, shown in (b), the curve for the PCA method is at least a factor of ten higher while the curve for the SHEEP method remains at about the same level as in Fig.~\ref{fig.AllanDev} (a).

Figure~\ref{fig.AllanDev} (c) shows the mean value of the measured rotation rate. The half-lengths of the error bar attached to each point are the value of the Allan deviation at one second. To guide to eye, in Fig.~\ref{fig.AllanDev} (c) we also show the applied rotation rate itself with a dot-dashed line. In the multiple-fringe regime, both the SHEEP and PCA methods return similar mean values and Allan deviations at one second. In the partial-fringe regime, the mean values of the PCA method deviate from the dot-dashed line and have larger Allan deviations. At zero applied rotation rate, the mean value of the SHEEP method is not exactly zero, due to the non-zero phase gradient offset discussed in Sec.~\ref{sec.TimeTrace}.

From Fig.~\ref{fig.AllanDev} we conclude again that the SHEEP method is more robust than the PCA method and has a wider dynamic range.

\section{Discussion}

\subsection{Measurable range of rotation rates}\label{sec.PiPuzzle}

In light-pulse atom interferometry, the rotation of the laboratory frame causes $\vec{k}_{\mathrm{eff}}$ to rotate such that atoms receive momentum kicks in slightly different directions from the sequential laser pulses, with the result that atomic spatial wavefunctions at the output of the interferometer do not completely overlap. This walk-off reduces the interferometer contrast at high rotation rates and sets a fundamental upper limit for the maximum rotation rate that can be measured. In our compact instrument, the finite-sized initial cloud and the small expansion factor also reduce the contrast of the spatial fringe pattern as the rotation rate increases \cite{Hoth.2016}, which also sets an upper limit on the measurable rotation rate. In Fig.~\ref{fig.TimeTrace}, the highest rotation rate of $5.1$~deg/s is not limited by the walk-off or reduced fringe contrast but simply by the upper limit of our ability to simulate rotation through the angular motion of the retroreflecting mirror. The lowest rotation rate of $0.045$~deg/s is limited by the electronic noise of the piezoelectric actuator electronics in the retroreflecting mirror.

The upper and lower limits on the measurable rotation rate also depend on technical factors such as the resolution of the analog-to-digital converter in the camera, the pixel size, the camera noise, and the system vibration noise. This experiment is built on a floating optical table without active vibration isolation. We use a relatively short Raman interrogation time $T_R=7.8$~ms to avoid the effects of vibration. This short $T_R$ is also compatible with the dimensions of the glass cell used in the experimental setup.

\subsection{Unambiguous rotation measurement}\label{sec.PiPuzzle}

In conventional atom interferometers, the rotation is measured from the rotation phase, which is typically derived from the population ratio via an inverse sinusoidal function. However, the inverse sine is ambiguous to an integer multiple of 2$\pi$, and therefore the measurement does not provide a unique value of the underlying rotation.

In contrast, PSI gives a unique value of the rotation. The interferometer rotation phase depends linearly on the atom velocity $v$ as $\phi_\Omega=2\vec{k}_\mathrm{eff}\cdot( \vec{\Omega} \times \vec{v} )T_R^2$, so the magnitude of the rotation depends on the quantity $d\phi_\Omega/dv=2k_\mathrm{eff}T_R^2\Omega$. The quantity $d\phi_\Omega/dv$ is inherently measured in PSI as a spatial phase gradient across the cold-atom cloud because of the position-velocity correlation of an expanding cloud. The spatial phase gradient is a one-to-one function of the rotation vector projected onto the image plane.

\section{Applications}
In light of the rotation dynamic range, unambiguity, and robustness provided by the SHEEP method, a PSI instrument may be operated in a free-running mode that returns rotation readings for inertial navigation applications. A PSI instrument may also be operated in a zero-fringe locking, closed-loop \cite{Cheinet.2006, Duan.2014, Gustavson.2000} mode that provides a rotation-free environment for another instrument. In the closed-loop mode, real-time phase maps generated by the SHEEP method are used as a servo input to an actuator that cancels the platform's rotation. In high-precision atom interferometer gravimeters, sources that generate spatial variation in the population distribution in the final cloud, such as the Earth's rotation \cite{Lan.2012}, the rotation of a moving platform, or the Raman laser wavefront \cite{Schkolnik.2015, Trimeche.2017}, can cause systematic errors in the acceleration measurement. The SHEEP method, combined with additional tools, may be used to characterize those spatially-dependent systematic effects \textit{in situ}.

\section{Conclusions}
We have demonstrated a new method which we termed ``Simple, High dynamic range, and Efficient Extraction of Phase map", or `` SHEEP''. The SHEEP method extracts the phase map of a point-source atom interferometer from four fringe images. We have compared the SHEEP method to a conventional fringe-analysis approach, in which images are processed with principal component analysis and subsequently fitted with parametric functions. We have demonstrated that the two methods are generally equivalent in the multiple-fringe regime (moderate and fast rotations); however, SHEEP is more robust in the partial-fringe regime (slow rotations).

The SHEEP method does not require prior knowledge of the fringe phase, fringe contrast, or the range of the rotation rate. These advantages benefit the experimental design, data acquisition, and analysis procedures in fieldable applications, as they considerably simplify the decision tree in instruments.

\section*{Acknowledgments}
We acknowledge valuable contributions from Gregory~W.~Hoth. We thank Kevin Coakley and Jolene Splett for helpful discussions in statistics. We thank William McGehee for valuable comments on our work. We thank Christopher Oates and Christopher Holloway for comments on the manuscript. A.~H. was supported for this work under an NRC Research Fellowship award at NIST. R.~B. was supported by the NIST Guest Researcher Program and the D\'el\'egation G\'en\'erale de l'Armement (DGA). Y.-J.~C. was supported under the financial assistance award 70NANB18H006 from U.S. Department of Commerce, National Institute of Standards and Technology. This work was funded by NIST, a U.S. government agency, and it is not subject to copyright.


\section*{Appendix}

\begin{figure}
\includegraphics[width=\columnwidth]{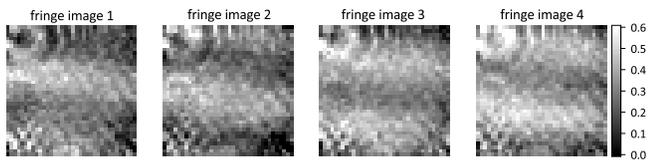}
\caption{Fringe images, plotted as the population ratio of the $F=2$ state. The imaging plane is perpendicular to the direction of the Raman laser beams. The fringe phases are $0$, $\pi/2$, $\pi$, and $3\pi/2$ relative to the unknown acceleration and Raman laser phase. All images have $36\times 36$~pixels and a physical size of $1.9\times 1.9$~mm$^2$.}\label{fig.Atom}
\end{figure}

Figure~\ref{fig.Atom} shows a set of four experimental fringe images with an applied rotation rate of 5.1~deg/s, which is the highest rotation rate studied in this work. We use commercial math software and its built-in functions to perform the 2D fit. The fit function uses a trust-region-reflective algorithm.

\subsection{SHEEP}
\begin{figure}
\includegraphics[width=\columnwidth]{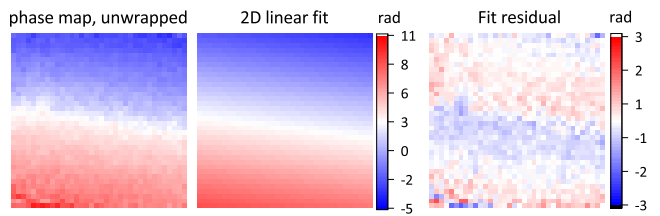}
\caption{Sample images for the SHEEP method. Left: phase map derived from the fringe images in Figure~\ref{fig.Atom}, with the inverse tangent function output unwrapped. Middle: 2D linear fitting to the phase map with the function in Equation~\ref{eqn.plane}. Right: fit residual. All images have $36\times 36$~pixels and a physical size of $1.9\times 1.9$~mm$^2$.}\label{fig.PM}
\end{figure}

For the SHEEP method, we do not perform any pre-processing on the fringe images. We convert a set of four fringe images into a phase map with Equation~\ref{eqn.atan} and then unwrap the inverse tangent output. We compute a set of initial fitting parameters and corresponding parameter boundaries from a basic initial statistical analysis of the phase map, and then apply a 2D linear fit using the model in Equations~\ref{eqn.plane} to the phase map. The phase map converted from the data in Fig.~\ref{fig.Atom} and the 2D linear fit are shown in Fig.~\ref{fig.PM}. We convert the fit parameters $k_x$ and $k_y$ and corresponding fit errors from each phase map into a measured rotation rate and the uncertainty.

\subsection{PCA}
\begin{figure}
\includegraphics[width=\columnwidth]{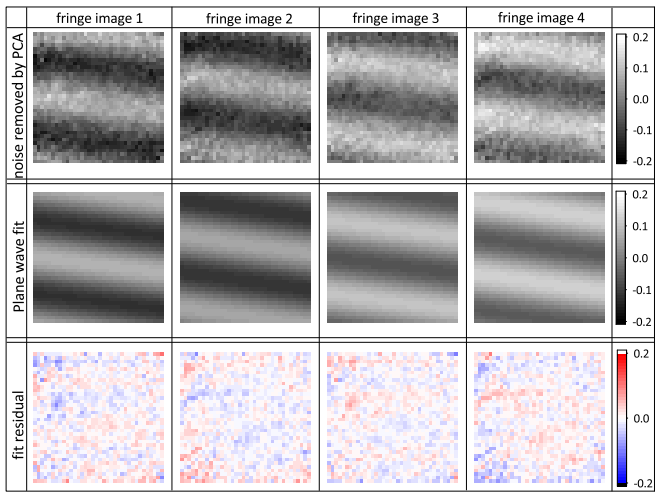}
\caption{Sample images for the PCA method. Top row: fringe images reconstructed by PCA. Middle row: 2D fitting to the reconstructed images with the plane wave function in Equation~\ref{eqn.planewave}. Bottom row: fit residual. All images have $36\times 36$~pixels and a physical size of $1.9\times 1.9$~mm$^2$.}\label{fig.PCA}
\end{figure}

For the PCA method, we first process the same set of four fringe images with PCA~\cite{gyro.2019}. A mean image is generated as the average of all four images. Each image is centered on this mean image by subtracting the mean image. The principal components of the four centered images are calculated. We use the leading two principal components to reconstruct the images without adding the mean image back to the reconstructed images. The reconstructed images for the data in Fig.~\ref{fig.Atom} are plotted in the top row of Fig.~\ref{fig.PCA}. The background noise is mostly removed in those processed images and the images are ready for 2D sinusoidal fitting.

A 2D sinusoidal fitting function repeatedly fits the same fringe image 50 times using the model in Equation~\ref{eqn.planewave}, with newly generated initial fit parameters each time. A uniform random number generator generates initial fit parameters between the upper and lower limits in Table~\ref{Tab.PlaneWave}. We use the optimized fit parameters from the trial that minimizes the residual. In this way we approximate a global search of parameters. We convert the fit parameters $k_x$ and $k_y$ and fit errors from each fringe image into a measured rotation rate and the uncertainty. We take the average of the rotation rates from the four fringe images and calculate the uncertainty in a typical way of the individual uncertainty divided by the square root of four.

\begin{table}
\begin{center}
\begin{tabular}{ cccccc }
\hline
& $k_x$ (rad/m) & $k_y$ (rad/m) & $\phi_0$ (rad) & $c$ & $g_0$ \\
\hline
lower & -7000 & -7000 & -3.14 & 0.01 & -0.5\\
upper & 7000 & 7000 & 3.14 & 0.99 & 0.5\\
\hline
\end{tabular}
\end{center}
\caption{Parameter limits for Equation~\ref{eqn.planewave} for the PCA method.}
\label{Tab.PlaneWave}
\end{table}

\newpage
\nocite{apsrev41Control}
\bibliographystyle{apsrev4-1}
\bibliography{bibliography}

\end{document}